# Topological Charge Quadrupole Protected by Spin-Orbit $U(1)$ Quasi-Symmetry in Antiferromagnet NdBiPt

Ao Zhang[1,4], Xiaobing Chen[2,1,4], Jiayu Li[1], Pengfei Liu[1], Yuntian Liu[1] and Qihang Liu[1,2,3,5,*]

[1]Department of Physics and Guangdong Basic Research Center of Excellence for Quantum Science, Southern University of Science and Technology, Shenzhen 518055, China

[2]Quantum Science Center of Guangdong–Hong Kong–Macao Greater Bay Area (Guangdong), Shenzhen 518045, China

[3]Guangdong Provincial Key Laboratory for Computational Science and Material Design, Southern University of Science and Technology, Shenzhen 518055, China

[4]These authors contributed equally to this work.

[5]Lead contact

[*]Correspondence: liuqh@sustech.edu.cn

**SUMMARY**

The interplay of symmetry and topology in crystal solids has given rise to various elementary excitations as quasiparticles. Among these, those with significant Berry-phase-related transport responses are of particular interest. In this study, we predict a type of quasiparticle called topological charge quadruple (TCQ), which is analogous to a charge quadrupole but consists of two closely packed pairs of Weyl points in momentum space, specifically in the half-Heusler antiferromagnet NdBiPt. Interestingly, the TCQ is protected by the spin-orbit $U(1)$ quasi-symmetry, rather than any exact crystallographic symmetries. This quasi-symmetry restricts the energy splitting induced by symmetry-lowering perturbations to a second-order effect. Furthermore, the closely located Berry curvature sources and sinks in the TCQ lead to a large Berry curvature dipole, resulting in a significant nonlinear Hall effect. Our work opens an avenue for designing unconventional quasiparticles utilizing quasi-symmetries and for developing materials with enhanced nonlinear responses.

**INTRODUCTION**

The investigation of topological semimetals and the discovery of gapless elementary excitations in these materials have been a prosperous area of research in condensed matter physics[1-4]. Various types of emergent quasiparticles, exemplified by Dirac[5-10], Weyl[11-19], and other fermions beyond them[20-23], have been predicted and observed in solids as low-energy fermionic excitations near the band crossings protected by symmetry and topology. The discovery of these quasiparticles has also led to the use of crystallographic symmetry-based scenario, i.e., space groups and magnetic space groups, to classify topological materials[24-33]. Recently, the classification of quasiparticles has been extended to include spin space groups, allowing for the treatment of magnetic materials with negligible spin-orbit coupling. This extension predicts various types of multifold band degeneracies and unconventional quasiparticles in both electronic and magnonic spectra[34-39]. Notably, the topological semimetals of particular interest are those hosting quasiparticles with observable transport responses. One such example is that magnetic Weyl semimetals exhibit the linear and nonlinear anomalous Hall effect[40-42] induced by Berry curvature. In contrast, Dirac semimetals, protected by $PT$-symmetry (where $P$ and $T$ denote space inversion and time reversal, respectively), exhibit a vanishing Berry curvature throughout momentum space, resulting in no related transport effects (Fig. 1a). Therefore, a long-sought goal is to identify ideal materials with topological quasiparticles close to the Fermi level that also exhibit significant transport properties.

In this work, we propose a type of quasiparticle, dubbed topological charge quadrupole (TCQ), in the noncentrosymmetric half-Heusler compound NdBiPt with a collinear antiferromagnetic (AFM) order. Analogous to the charge quadrupole, the TCQ consists of four closely located Weyl points in momentum space, with two possessing a topological charge of +1 and the other two having a topological charge of -1, as shown in Fig. 1b. Interestingly, such a quasiparticle complex is not protected by exact symmetry but rather by a novel spin-orbit $U(1)$ quasi-symmetry[43,44]. While the electronic structure of NdBiPt closely resembles a Dirac semimetal, the four nearly degenerate Weyl points manifest Berry curvature sources and sinks within a small region of momentum space, leading to a strong Berry curvature dipole (BCD) and nonlinear Hall effects (Fig. 1b). The TCQ of NdBiPt results in a significant BCD peak located only 5.8 meV below the Fermi level, with a peak value comparable to those of the $T_d$ phase of WTe$_2$[45] and collinear AFM CuMnSb[42]. Furthermore, the strong dependence of the BCD on the Néel vector can be used for detecting the Néel vector itself. Our work not only reveals the important role of hidden symmetry in predicting new quasiparticles, but also provides an approach for designing significant quantum geometric effects in unconventional AFM materials.

**RESULTS**

*Topological charge quadrupole in NdBiPt*

Recent neutron diffraction measurements show that NdBiPt accommodates an A-type collinear AFM configuration[46] along the [001] direction of the nonmagnetic cubic conventional cell below the Néel temperature $T_N = 2.18\ K$, resulting in the tetragonal magnetic unit cell with lattice constants $a = b = 4.78$ Å and $c = 6.76$ Å (Fig. 2a). The magnetic geometry and the SOC-free electronic structure are fully described by its spin space group $G_S = P\ ^1\bar{4}^1m^12^{-1}(1/2\ 1/2\ 1/2)^{\infty m}1$ (No. 115.119.2.1), which is identified by the online program FINDSPINGROUP[47,48]. Especially, the collinear magnetic order is described by the spin-only group $G_{SO} = ^{\infty m}1 = SO(2) \rtimes Z_2^K$ (only contains pure spin operations), where $SO(2)$ and $Z_2^K$ represent the continuous spin rotations along the collinear spin axis and the mirror symmetry $TU$ in spin space ($U$ is the twofold rotation along any axis perpendicular to the collinear spin axis), respectively (see Supplemental Methods A for detailed symmetry analysis). The DFT-calculated band structure without SOC is shown in Fig. 2c (see methods). Two prominent features are observed. The first one is the double degeneracy throughout the whole Brillouin zone protected by the $U\tau$ and $SO(2)$ symmetries[39,47], where $\tau = (1/2, 1/2, 1/2)$ is the fractional translation symmetry that connects the two opposite-spin sublattices. Notably, $U\tau$-enforced double degeneracy provides emergent quasiparticles with doubled topological charges from two $P$-broken spin channels[37,38], which differs from compensated ones enforced by $PT$ symmetry[39]. Secondly, a four-fold degenerate $\beta$ band (orange line) intersects with a two-fold degenerate $\alpha$ band (green line), forming a six-fold degenerate quasiparticle at the $Q(0, 0, 0.006\frac{2\pi}{c})$ point. Such a $\beta$ band comprises a Dirac nodal line along $-Q - \Gamma - Q$, as shown in Fig. 2d, while the nodal line splits off this high-symmetry line (Fig. S1).

Since NdBiPt is composed of heavy elements, the corresponding electronic structure with SOC is expected to exhibit significant SOC-induced band splitting as shown in Fig. 2e. The electronic structure with SOC is dictated by its magnetic space group $G_M = P_I\bar{4}n2$ (No. 118.314). In the presence of SOC, both $U\tau$ and $SO(2)$ symmetries are broken, leading to the spin-polarized non-degenerate bands. However, the bands along the $\Gamma - Z$ direction maintain twofold degeneracy, which is protected by two-fold rotation $\{2_{001}|0\}$ and glide mirror $\{m_{010}|\tau\}$ (see Supplemental Methods A). Additionally, the fourfold degenerate $\beta$ band splits into $\beta_1$ and $\beta_2$. Surprisingly, while the SOC gap between $\alpha$ and $\beta_2$ ($\Delta E_{\alpha\beta_2}$) at the $Q$ point remarkably reaches 672.4 meV (indicated by the black arrow), an unexpectedly small SOC gap $\Delta E_{\alpha\beta_1} = 9.2$ meV emerges between $\alpha$ and $\beta_1$. To illustrate the evolution of the SOC effect, we present the electronic structures with varying SOC strength $\lambda$ in Fig. S2. Specifically, as $\lambda$ increases, $\Delta E_{\alpha\beta_2}$ exhibits a significant linear increase from 0 to 672.4 meV, while $\Delta E_{\alpha\beta_1}$ increases gradually and eventually reaches 9.2 meV (Fig. 2f). The substantial difference between $\Delta E_{\alpha\beta_1}$ and $\Delta E_{\alpha\beta_2}$ cannot be explained by traditional group representation theory because $\alpha$, $\beta_1$, and $\beta_2$ bands share the same two-dimensional irreducible representation $\bar{Q}_5$ of the little group $G_M^Q = \bar{4}'m2'$ at the $Q$ point.

We next look into the fine electronic structure within the tiny gap of $\Delta E_{\alpha\beta_1}$, where multiple Weyl points reside. In the noncentrosymmetric $T\tau$-AFM NdBiPt, each Weyl point has a $T\tau$ partner of the same chirality. Meanwhile, two glide mirror symmetries, $\{m_{100}|\tau\}$ and $\{m_{010}|\tau\}$, connect Weyl points of opposite chirality. Consequently, an AFM Weyl semimetal with eight Weyl nodes is confirmed by DFT calculations, as shown in Fig. 3a. Taking a close look off the $Q$ point, we have uncovered two pairs of Weyl points with charge $\pm 1$ (Fig. S3). Due to the tiny gap at $Q$, these Weyl points form a square with a side length of 0.01Å$^{-1}$, which is only 1/131 of the in-plane reciprocal lattice parameter. Therefore, such a closely packed configuration of Weyl points forms a topological charge quadrupole (TCQ) in momentum space. Because of the $T\{2_{001}|\tau\}$ symmetry, the chirality distribution around $Q$ and $-Q$ points are the same, as shown in Fig. 3a. Therefore, in the (001) surface Brillouin zone, the Weyl points are projected pairwise on four different projected positions $W_i$ ($i = 1, 2, 3, 4$) with an effective topological charge of $\pm 2$. Such a distribution of Weyl points results in two overlapping Fermi rings composed of four individual Fermi arcs, and further forms two closed Fermi rings due to band hybridization (see the right panel of Fig. 3a). We calculate the Fermi arc surface states on the (001) surface, where the two closed Fermi ring surface states are presented in Fig. 3b. The smaller Fermi ring connecting the TCQ is further shown in Fig. 3c.

*Spin-orbit U(1) quasi-symmetry*

The key feature of the TCQ is the near degenerate Weyl points, which cannot be well explained by any exact symmetries. However, recent studies showed that the classical group representation theory could be extended to address the issue of near degeneracy by using quasi-symmetry[43,44,49]. Different from the general concept of approximate symmetry[50], quasi-symmetry refers to the hidden symmetry within a degenerate orbital subspace under an unperturbed Hamiltonian $H^0$. This limits the occurrence of the symmetry-lowering term, denoted as $H'$, to only second-order effects. While the previous studies have primarily focused on nonmagnetic systems, we next demonstrate that the formation of the TCQ in AFM NdBiPt can be attributed to a type of spin-orbit $U(1)$ quasi-symmetry. This symmetry results in a partially lifted band degeneracy under the first-order SOC effect, and thus a tiny gap when involving the second-order SOC effect at the $Q$ point.

By analyzing the DFT-calculated orbital projection of the bands along the $\Gamma - Z$ direction (Figs. S4-S6 and Table S2), we can effectively write the $\alpha$, $\beta_1$, and $\beta_2$ bands as:

$$\begin{cases} \alpha = (|l_0, \uparrow\rangle + |l_0, \downarrow\rangle)/\sqrt{2} \\ \beta_1 = (|l_+, \uparrow\rangle + |l_-, \downarrow\rangle)/\sqrt{2} \\ \beta_2 = (|l_-, \uparrow\rangle + |l_+, \downarrow\rangle)/\sqrt{2} \end{cases} \qquad \text{(Equation 1)}$$

where $l_0$ and $l_\pm$ can be expressed as linear combinations of the atomic orbitals, designated as $|l_0\rangle = \theta|s\rangle + \gamma|p_z\rangle$ and $|l_\pm\rangle = \frac{1}{\sqrt{2}}(|u_x\rangle \pm i|u_y\rangle)$, where $|u_{x/y}\rangle = \eta|p_{x/y}\rangle + \delta|d_{xz/yz}\rangle$. In the absence of SOC, the symmetry of the Hamiltonian $H_A^0 = H_k + H_p + H_{mag}$, which includes the kinetic term $H_k$, the potential term $H_p$ and the magnetic term $H_{mag}$, can be described by $G_{H_A^0} = G_S$. The doubly degenerate band $\alpha$ and the four-fold degenerate band $\beta$ with irreps $Q_{1,A}^{1/2}(2)$ and $Q_{3,A}^{1/2}Q_{4,A}^{1/2}(4)$ of the little group $G_{H_A^0}^Q = {}^m\bar{4}^1m^m2^{\infty 2}1$ intersect at the $Q$ point as shown in Fig. 4a. When including SOC, $\beta$ band splits into two doubly degenerate bands $\beta_1$ and $\beta_2$. The band crossing between $\alpha$ and $\beta_1$ is gapped (and so is that between $\alpha$ and $\beta_2$), because $\alpha$, $\beta_1$, and $\beta_2$

bands share the same magnetic group representation $\overline{Q}_5(2)$ (Fig. 4d). We next use the quasi-symmetry theory to elucidate the remarkable difference of two SOC-induced gaps at the $Q$ point, $\Delta E_{\alpha\beta_1}$, and $\Delta E_{\alpha\beta_2}$.

Note that the difference between $\Delta E_{\alpha\beta_1}$ and $\Delta E_{\alpha\beta_2}$ cannot be captured using any group extended by $G^Q_{H^0_A}$, which failed in separating bands $\beta_1$ and $\beta_2$ (see Supplemental Methods B). Interestingly, we find that the $\lambda L_z S_z$ term does not contribute to the two SOC gaps but splits the $\beta$ band into bands $\beta_1$ and $\beta_2$ (see Supplemental Methods C). Therefore, we consider the unperturbed Hamiltonian $H^0_B = H^0_A + \lambda L_z S_z$ and the perturbed Hamiltonian $H'_B = \frac{\lambda}{2}(L_+ S_- + L_- S_+)$. The introduction of $\lambda L_z S_z$ breaks the little group $G^Q_{H^0_A}$ into $G^Q_{H^0_B} = \overline{1}\overline{4}^2 m^m 2^\infty 1$, thus providing an adequate starting point of three doubly degenerate bands (Fig. 4b). Then we find that the spin-orbit $U(1)$ symmetry $P_q = \{U_z(\theta)\|R_z(\theta)\}$[51,52], which represents the simultaneously continuous rotation along the $z$-direction in both spin and lattice space, is not in $G^Q_{H^0_B}$, but commutes with $H'_B$. It transforms the matrix element $\langle\alpha|H'_B|\beta_1\rangle$ into an additional phase factor, i.e., $\langle\alpha|H'_B|\beta_1\rangle \xrightarrow{P_q} e^{i\omega(P_q)}\langle\alpha|H'_B|\beta_1\rangle$, thereby enforcing the first-order SOC effect between $\alpha$ and $\beta_1$ bands to be zero[44]. In contrast, two elements of $\langle\alpha|H'_B|\beta_2\rangle$ are transformed as an identity representation (see Supplemental Methods D), resulting in a relatively larger gap as shown in Fig. 4c. Consequently, the first-order SOC-induced gap at $Q$ is zero between $\alpha$ and $\beta_1$ bands, leading to a small second-order gap of 9.2 meV. In contrast, the band crossing of $\alpha$ and $\beta_2$ bands is lifted by first-order SOC with 672.4 meV (Fig. 4d).

To validate the above quasi-symmetry analysis, we constructed a $k \cdot p$ model near the $\Gamma$ point (see Supplemental Methods E). The distinction between two SOC gaps can be estimated utilizing the first-order SOC Hamiltonian $H^{(1)}_{soc} = \lambda L \cdot S$ (See Eq. S27),

$$\Delta E^{(1)}_{\alpha\beta_1} \sim \langle\alpha|H^{(1)}_{soc}|\beta_1\rangle = \lambda/2(h_{15} + h_{25} + h_{16} + h_{26}) = 0 \quad \text{(Equation 2)}$$

$$\Delta E^{(1)}_{\alpha\beta_2} \sim \langle\alpha|H^{(1)}_{soc}|\beta_2\rangle = \lambda/2(h_{13} + h_{23} + h_{14} + h_{24}) = \sqrt{2}\lambda/2. \quad \text{(Equation 3)}$$

Eqs. (2) and (3) show that the first-order SOC effect is prohibited for $\Delta E_{\alpha\beta_1}$ but is allowed for $\Delta E_{\alpha\beta_2}$, consistent with the framework of quasi-symmetry analysis. Therefore, we demonstrate that the TCQ, two closely packed pairs of Weyl points within the tiny gap at the $Q$ point, is protected by spin-orbit $U(1)$ quasi-symmetry.

**DISCUSSION**

The TCQ gathers the sources and sinks of Berry curvature in a very small region in momentum space, leading to a remarkable BCD. Thus, an enhanced nonlinear Hall effect is expected if the TCQ emerges around the Fermi level, as shown in Fig. 1b. In $T\tau$-AFM NdBiPt, there is only one independent element $D_{xy} = D_{yx}$ of the BCD tensor $\boldsymbol{D}$ (see Eq. S2), where $D_{xy} = \int_{BZ} dk \sum_n f_0(\varepsilon_{nk})(\frac{\partial}{\partial k_x}\Omega_y)$. The distribution of the Berry curvature $\Omega_y$ on the $k_x - k_y$ plane is shown in Fig. 4e. We find that $\Omega_y$ is mainly concentrated around the TCQ. Due to the presence of two glide mirrors $\{m_{100}|\tau\}$ and $\{m_{010}|\tau\}$, $\Omega_y$ is antisymmetric along the $k_x$ direction and symmetric along the $k_y$ direction. The gapped $Q$ point (black triangles in Fig. 4e) is located between the peaks of the positive and negative Berry curvature around the TCQ. Furthermore, the distribution of $\Omega_y$ in the vicinity of the Fermi surface is non-uniform due to the tilting of the Weyl cones. Overall, such a distribution pattern of the Berry curvature gives rise to a large BCD (see Figs. S7-S8).

The BCD density $d_{xy}$ along the high-symmetry paths is shown in Fig. 4f, indicating that the origin of $d_{xy}$ comes entirely from the contribution of the TCQ around the $Q$ point. Since the BCD is the integral of $d_{xy}$, it is consistent with our calculation, which shows that the BCD exhibits a peak at the position of the TCQ, as shown in Fig. 4g. The large BCD value approaches 0.02 when the chemical potential is positioned at the TCQ (indicated by the black dashed line in Fig. 4g). This value is comparable to the previous study on CuMnSb[42], where the BCD peak is located at about 100 meV above the Fermi level. In comparison, due to the clean Fermi surface composed of the TCQ, the BCD peak of NdBiPt is located only 5.8 meV below the Fermi level, making it favorable for experimental detection. Moreover, we expect that the TCQ can also be utilized to design a substantial circular photogalvanic effect, which shares close similarities with the Berry curvature dipole in terms of symmetry constraints.

In summary, based on the spin group analysis, we extend the quasi-symmetry theory to elucidate the formation of the TCQ quasiparticle, which is the analogous to charge quadrupole but composed of two closely packed pairs of Weyl points in momentum space. Our DFT calculations show that the noncentrosymmetric half-Heusler compound NdBiPt, with a collinear AFM configuration, is an ideal material candidate, manifesting a clean Fermi surface with TCQ. The structure of the TCQ is protected by mirror symmetry and, more importantly, the spin-orbit $U(1)$ quasi-symmetry, which eliminates the first-order SOC gap and leads to a significant nonlinear Hall effect induced by BCD. Overall, our research provides an avenue for designing unconventional quasiparticles through quasi-symmetries and also for designing material candidates with large nonlinear physical responses. Furthermore, the exotic transport properties induced by the TCQ can also facilitate the detection of the Néel vector, providing additional possibilities for antiferromagnetic spintronics.

**METHODS**

The first-principles calculations utilized the projector-augmented-wave (PAW) method[53], within the Vienna ab initio simulation package (VASP)[54]. The exchange and correlation effects were treated by the generalized gradient approximation (GGA) with the Perdew-Burke-Ernzerhof (PBE) formalism[55]. An energy cut-off of 253 eV was employed for the calculations. The whole Brillouin zone was sampled by an 11×11×8 Monkhorst-Pack grid[56] for all cells. Due to the local magnetic moments contributed by $f$ electrons in Nd atoms, the GGA+$U$ approach[57] within the Dudarev scheme[58] was applied. In our calculations, the topological charge quadrupole of NdBiPt was relatively insensitive to the choice of $U_{Nd}$ values (see Fig. S9). For the convenience of discussion, we elaborated on the details using $U_{Nd}$ = 4 eV as an example. A tight-binding Hamiltonian was obtained based on maximally localized Wannier functions of Nd $d$ and $f$ orbitals, Bi $p$ orbitals, and Pt $s$ and $d$ orbitals. The maximally localized Wannier function was constructed using the WANNIER90 package[59,60] and the WannierTools package[61] was utilized to calculate the surface states and the position of Weyl points. The WANNIERBERRI package[62] was used to calculate the Berry curvature dipole, where a 500 × 500 × 500 $k$-point mesh was used to achieve the convergence. The computation of the Berry curvature dipole is performed within the Fermi sea integral method, which is given as $D_{ab} = \int_{BZ} dk \sum_n f_0(\varepsilon_{nk})(\frac{\partial}{\partial k_a}\Omega^b_{nk})$, where $D_{ab}$ is the Berry curvature dipole, $f_0(\varepsilon_{nk})$ is the Fermi-Dirac distribution, and $\Omega^b_{nk}$ denotes the Berry curvature. Further details regarding the methods can be found in the Supplemental Methods.

## RESOURCE AVAILABILITY

### *Lead contact*

Requests for additional information and resources should be directed to the lead contact, Qihang Liu (liuqh@sustech.edu.cn), who will address these inquiries.

### *Materials availability*

This study did not generate new materials.

### *Data and code availability*

All data needed to evaluate the conclusions in the paper are present in the paper and/or the Supplementary Materials. Additional data related to this paper may be requested from the authors.

## ACKNOWLEDGEMENTS

This work was supported by the National Key R&D Program of China under Grant Nos. 2019YFA0704900 and 2020YFA0308900, the National Natural Science Foundation of China under Grant No. 12274194, Guangdong Provincial Key Laboratory for Computational Science and Material Design under Grant No. 2019B030301001, Innovative Team of General Higher Educational Institutes in Guangdong Province (No. 2020KCXTD001), Shenzhen Science and Technology Program (Grants No. RCJC20221008092722009 and No. 20231117091158001) and Center for Computational Science and Engineering of Southern University of Science and Technology.

## AUTHOR CONTRIBUTIONS

Q.L. conceived and designed the work. A.Z. and X.C. performed calculations. A.Z., X.C., J.L., P.L., and Y.L. analyzed the results. A.Z., X.C., and Q.L. wrote the manuscript. All authors discussed the results and commented on the manuscript.

## DECLARATION OF INTERESTS

The authors declare that they have no competing interests.

## FIGURE TITLES AND LEGENDS

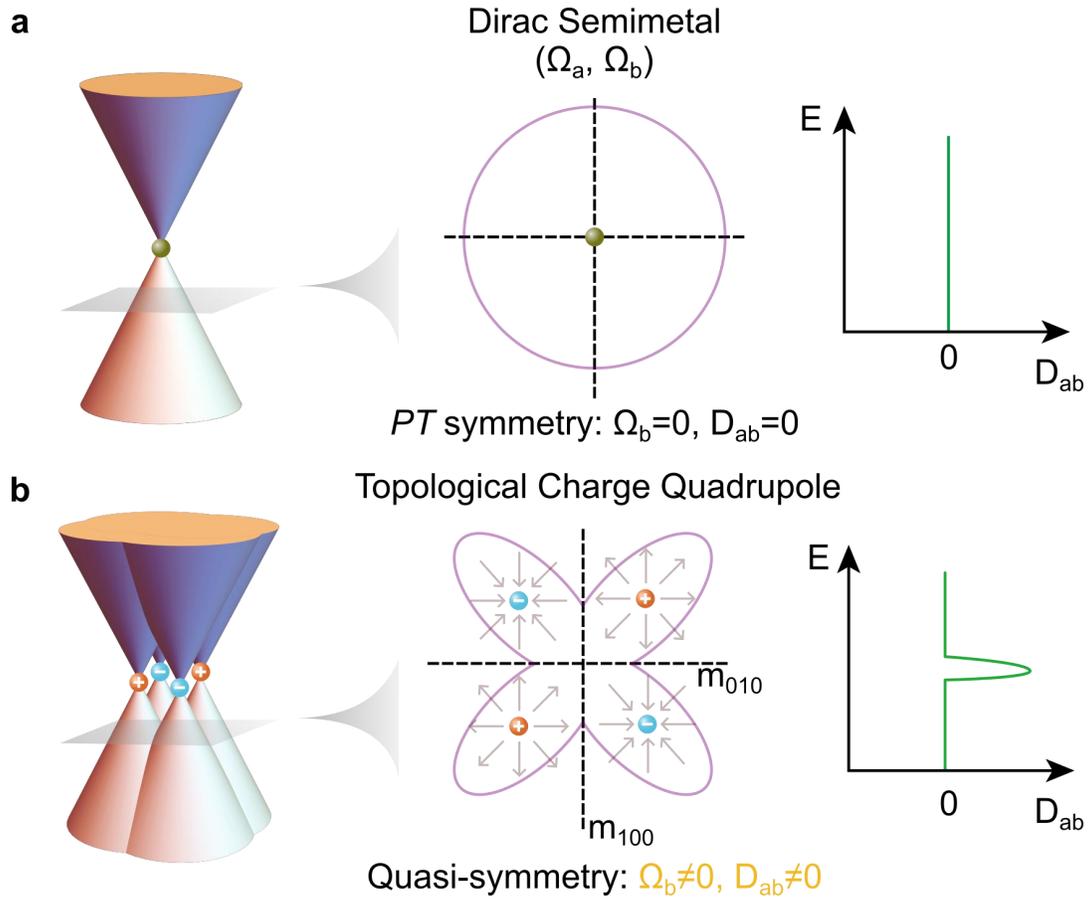

**Fig. 1. Schematics of the Dirac fermion and the topological charge quadrupole (TCQ). a** A Dirac semimetal protected by *PT* symmetry exhibits zero Berry curvature ($\Omega_b$) and zero Berry curvature dipole ($D_{ab}$). **b** A topological charge quadrupole is protected by the quasi-symmetry, resulting in a significant Berry curvature and Berry curvature dipole. The orange and blue spheres represent Weyl points with the monopole charge +1 and -1, respectively. The middle panels in **a-b** represent Fermi surfaces and Berry curvatures of the Dirac cone and TCQ, respectively.

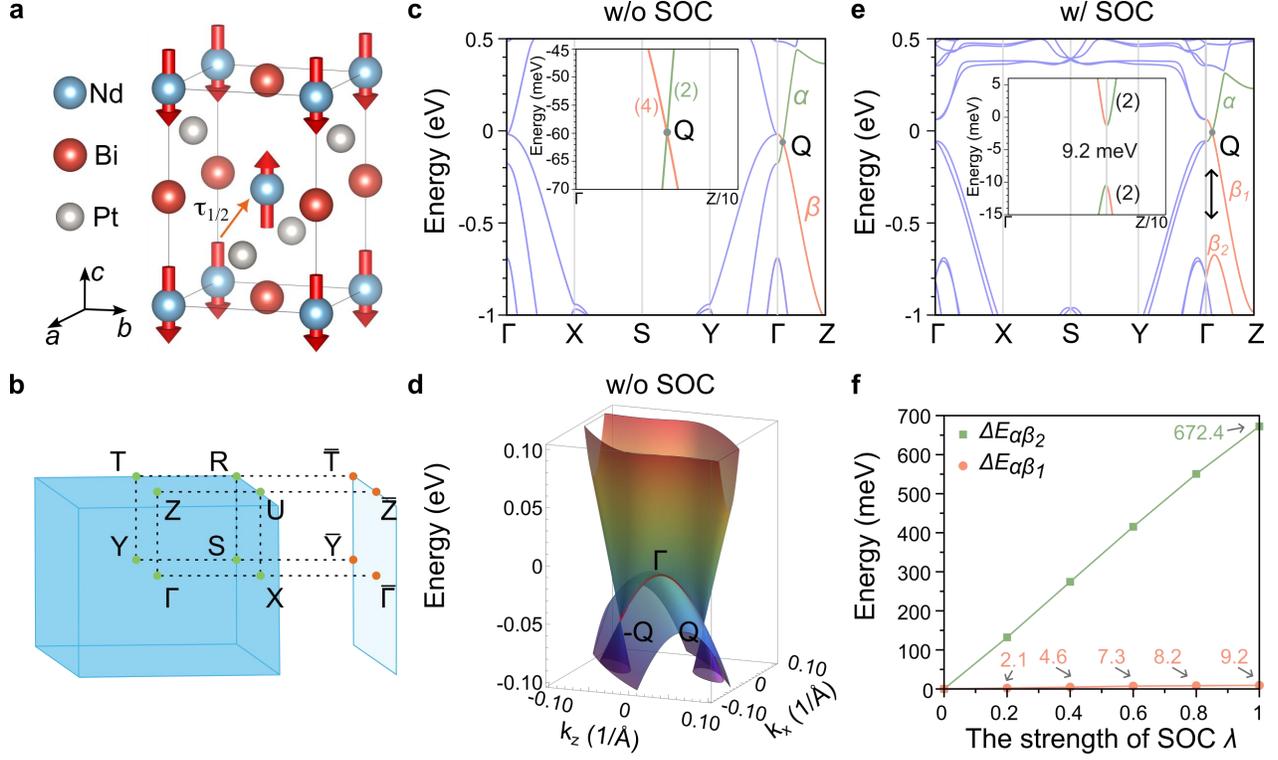

**Fig. 2. Magnetic structure, Brillouin zones, and electronic structures of NdBiPt. a** The magnetic structure of collinear AFM NdBiPt. **b** Bulk and surface Brillouin zones (BZs) of NdBiPt. **c** The electronic structure of NdBiPt without spin-orbit coupling (SOC). The inset in **c** shows the degeneracy between $\alpha$ and $\beta$ bands. **d** Three-dimensional electronic structure of NdBiPt. The red line denotes the Dirac nodal line within the $k_z - k_x$ plane without SOC. **e** The electronic structure of NdBiPt with SOC. The inset in **e** shows the tiny gap between $\alpha$ and $\beta_1$ bands. The green and orange bands denote the $\alpha$ and $\beta_1/\beta_2$ bands, respectively. The black arrow indicates the gap $\Delta E_{\alpha\beta_2}$ between $\alpha$ and $\beta_2$ at the $Q$ (0, 0, 0.006) point (fractional coordinates). The Fermi level is set to zero. **f** The band gaps $\Delta E_{\alpha\beta_1}$ and $\Delta E_{\alpha\beta_2}$ as a function of the spin-orbit coupling strength $\lambda$.

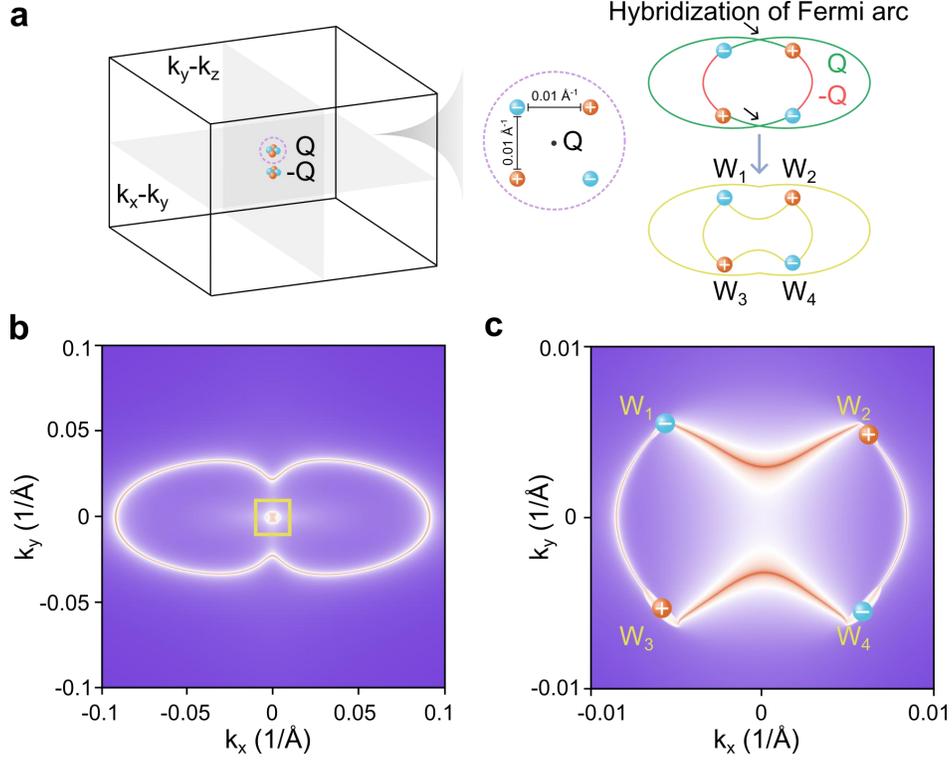

**Fig. 3. The distribution and Fermi arcs of TCQ. a** The distribution of topological charge quadrupole in the Brillouin zone of NdBiPt. The orange and blue spheres represent Weyl points with the monopole charge +1 and -1, respectively. The right panel of **a** shows a schematic diagram of two closed Fermi rings formed by the hybridization of Fermi arcs. The green and red lines represent the Fermi arcs located at $Q$ and $-Q$, respectively. **b** The isoenergy surface states enclosing topological charge quadrupole on the $k_x - k_y$ plane at 5.8 meV below the Fermi energy. **c** An enlarged view of the yellow square is marked in **b**.

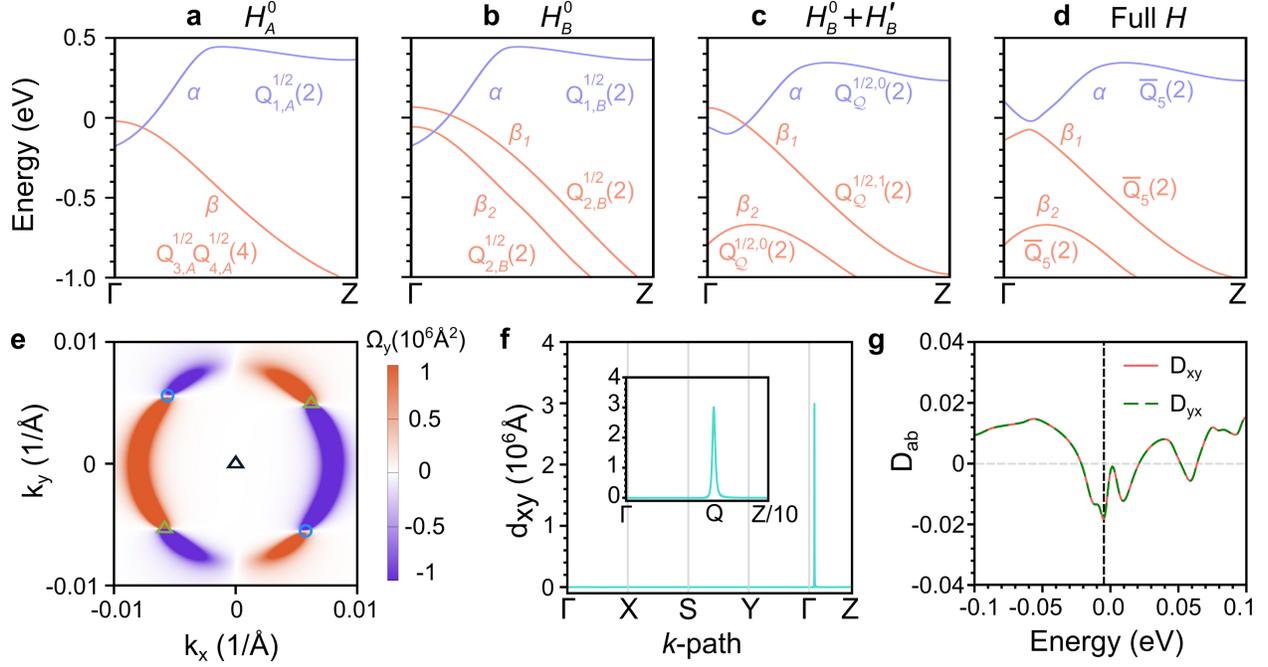

**Fig. 4. Quasi-symmetry and the nonlinear Hall effect in NdBiPt.** Evolution of bands $\alpha$, $\beta_1$, and $\beta_2$ for NdBiPt when considering different Hamiltonians: **a** the SOC-free Hamiltonian $H_A^0$, **b** the Hamiltonian that includes diagonal first-order SOC terms $H_B^0 = H_A^0 + \lambda L_z S_z$, **c** the Hamiltonian that includes off-diagonal first-order SOC terms $H_B^0 + H_B'$, where $H_B' = \frac{\lambda}{2}(L_+S_- + L_-S_+)$, and **d** the Hamiltonian $H$ that considers the full SOC effects. **e** Distribution of the Berry curvature within the $k_x - k_y$ plane. **f** BCD density $d_{xy}$ along the high-symmetry line. The inset in **f** denotes $d_{xy}$ along the $\Gamma - Z/10$. **g** Berry curvature dipole $D_{ab}$ as a function of energy. The black dashed line represents the energy position of the $Q$ point, which is 5.8 meV below the Fermi energy.